\documentclass[aip,12pt,preprint,amsmath,amssymb,groupedaddress]{revtex4-1}

\usepackage{graphicx}
\usepackage{amsmath,amssymb,latexsym}

\usepackage{times}
\usepackage{mathptmx}

\newcommand{\sol}{SOL}

\newcommand{\wt}[1]{{\widetilde#1}}

\newcommand{\Ip}{\ensuremath{I_\text{p}}}

\newcommand{\Figref}[1]{Fig.~\ref{#1}}
\newcommand{\Figsref}[1]{Figs.~\ref{#1}}

\newcommand{\Tabref}[1]{Tab.~\ref{#1}}

\newcommand{\CJP}{\textit{Czech.\ J.\ Phys.}}
\newcommand{\JNM}{\textit{J.~Nuclear Mater.}}

\newcommand{\NF}{\textit{Nucl.\ Fusion}}
\newcommand{\NME}{\textit{Nucl.\ Mater.\ Energy}}

\newcommand{\PPCF}{\textit{Plasma Phys.\ Contr.\ Fusion}}
\newcommand{\PFR}{\textit{Plasma Fusion Res.}}

\newcommand{\PP}{\textit{Phys.\ Plasmas}}
\newcommand{\PRE}{\textit{Phys.\ Rev.~E}}
\newcommand{\PS}{\textit{Phys.\ Scripta}}
\newcommand{\RSI}{\textit{Rev.\ Sci.\ Instr.}}

\newcommand{\PRL}{\textit{Phys.~Rev.\ Lett.}}

\begin{document}

\title{Intermittent fluctuations in the Alcator C-Mod scrape-off layer for ohmic and high confinement mode plasmas}
\date{\today}

\author{O.~E.~Garcia}
\email{odd.erik.garcia@uit.no. Invited speaker 59th APS DPP 2017}
\author{R.~Kube}
\author{A.~Theodorsen}
\affiliation{Department of Physics and Technology, UiT The Arctic University of Norway, N-9037 Troms{\o}, Norway}
\author{B.~La{B}ombard}
\author{J.~L.~Terry}
\affiliation{MIT Plasma Science and Fusion Center, Cambridge 02139, MA, USA}

\begin{abstract}
Plasma fluctuations in the scrape-off layer of the Alcator C-Mod tokamak in ohmic and high confinement modes have been analyzed using gas puff imaging data. In all cases investigated, the time series of emission from a single spatially-resolved view into the gas puff are dominated by large-amplitude bursts, attributed to blob-like filament structures moving radially outwards and poloidally. There is a remarkable similarity of the fluctuation statistics in ohmic plasmas and in edge localized mode-free and enhanced D-alpha high confinement mode plasmas. Conditionally averaged wave forms have a two-sided exponential shape with comparable temporal scales and asymmetry, while the burst amplitudes and the waiting times between them are exponentially distributed. The probability density functions and the frequency power spectral densities are self-similar for all these confinement modes. These results are strong evidence in support of a stochastic model describing the plasma fluctuations in the scrape-off layer as a super-position of uncorrelated exponential pulses. Predictions of this model are in excellent agreement with experimental measurements in both ohmic and high confinement mode plasmas. The stochastic model thus provides a valuable tool for predicting fluctuation-induced plasma--wall interactions in magnetically confined fusion plasmas.
\end{abstract}

\maketitle

\section{Introduction}

The life-time of plasma facing components at the outboard mid-plane region in the next generation magnetic confinement experiments and future fusion reactors is likely to be limited by enhanced erosion rates due to radial motion of blob-like filament structures through the scrape-off layer (SOL).\cite{labombard-2001,pigarov,pitts,lipschultz,whyte,dm-pwi,birkenmeier,marandet,brooks,dmz} It is therefore of great interest to elucidate the statistical properties of the intermittent fluctuations in the boundary region for reactor relevant conditions and plasma parameters.\cite{boedo1,rudakov,antar,boedo2,carreras,graves,horacek,rudakov-nf,garcia-tcv-fec,garcia-tcv-eps,xu,garcia-pfr,horacek-asdex,militello,carralero-nf,zweben-2015,carralero-prl,zweben-2016,boedo3,walkden} In particular, the rate of erosion will depend on the amplitude of the structures, the duration of the events, and their frequency of occurence.\cite{garcia-acm-psi,garcia-acm-aps,theodorsen-nf,ghp,kube-ppcf,garcia-nme,theodorsen-ppcf,militello-nf-2016,militello-ppcf-2016,walkden-ppcf-2017} If there are universal statistical properties of the fluctuations, it may be possible to give reliable predictions of fluctuation-induced plasma--wall interactions by use of phenomenological statistical models.\cite{garcia-acm-psi,garcia-acm-aps,theodorsen-nf,ghp,kube-ppcf,garcia-nme,theodorsen-ppcf,militello-nf-2016,militello-ppcf-2016,walkden-ppcf-2017,garcia-prl,kg,tg,gktp,gt,tgr,tg-pdf,tg-x,marandet-2016,marandet-2017}

In order to identify the statistical properties of the fluctuations in the SOL, exceptionally long measurement data time series under stationary plasma conditions in ohmic and low confinement mode (L-mode) plasmas have previously been carefully analyzed.\cite{garcia-acm-psi,garcia-acm-aps,theodorsen-nf,ghp,kube-ppcf,garcia-nme,theodorsen-ppcf} It has been demonstrated that the fluctuations are strongly intermittent in the far-SOL with an exponential tail in the probability density function (PDF) for large fluctuation amplitudes. In ohmic plasmas, the frequency power spectral density has been shown to be similar for all radial positions in the SOL and line-averaged densities.\cite{theodorsen-nf} Based on this, a stochastic model of the plasma fluctuations has been developed and its underlying assumptions and predictions are found to compare favorably with experimental measurements in ohmic and L-mode plasmas.\cite{garcia-acm-psi,garcia-acm-aps,theodorsen-nf,ghp,kube-ppcf,garcia-nme,theodorsen-ppcf,militello-nf-2016,militello-ppcf-2016,walkden-ppcf-2017}

In this work, it is demonstrated for the first time that the plasma fluctuations in the SOL of Alcator C-Mod have the same statistical properties in ohmic and high confinement mode (H-mode) plasmas. The latter includes both an ELM-free H-mode and an enhanced D-alpha (EDA) H-mode confinement regimes. In particular, it is shown that large-amplitude bursts in the SOL data time series have an exponential wave form with constant duration. Both the peak amplitudes of these bursts and the waiting times between them are exponentially distributed. Moreover, the frequency power spectral densities in the SOL have a self-similar shape for both ohmic and H-mode plasma states. This gives further evidence for universality of plasma fluctuations in the SOL of magnetically confined plasmas and supports the stochastic model.

\section{Experimental setup}

Alcator C-Mod is a compact, high-field tokamak with major radius $R_0=0.68\,\text{m}$ and minor radius $a=0.21\,\text{m}$.\cite{hutchinson,greenwald-2013,greenwald-2014} All experiments analyzed here were deuterium fuelled plasmas in a lower single null divertor configuration. There is a wide range of plasma operation conditions available for the Alcator C-Mod tokamak. In the case of strong auxiliary ion cyclotron range of frequencies (ICRF) heating, there are two different types of H-modes on Alcator C-Mod without edge localized modes (ELMs). The most common is the so-called enhanced D-alpha (EDA) H-mode, which is a steady mode of operation with an edge transport barrier. Enhanced particle transport has been correlated with a quasi-coherent mode
(QCM) observed in the particle density and magnetic fluctuations at frequencies between $50$ and $200\,\text{kHz}$.\cite{terry-2005,cziegler-2010,labombard-qcm}
This mode is believed to prevent impurities from accumulating in the core, resulting in a steady state EDA H-mode without ELMs.

Another type of H-mode on Alcator C-Mod is the so-called ELM-free H-Mode. In this case there is a strong particle and heat transport barrier but a lack of macroscopic instabilities of the edge pedestal. This results in an accumulation of impurities in the core, which eventually causes a radiative collapse of the plasma. Both the plasma and impurity densities increase monotonically during these ELM-free H-modes. ELM-free H-modes are therefore inherently transient in nature. However, from the point of view of the far-SOL turbulence properties, this is an interesting mode of confinement due to the absence of a transport regulator in the edge region. Finally, it is to be mentioned that the H-mode data sets analyzed here have been carefully chosen such that the GPI measurements are not influenced by the strong electric fields from the ICRF wave antennas. When the GPI field of view is magnetically mapped along field lines to the antenna, there are significant changes in the dynamics of blob-like filament structures,\cite{cziegler-2012} and the fluctuations are found to be near normally distributed throughout the SOL. Such interactions are beyond the scope of this study.

The GPI diagnostic on Alcator C-Mod consists of a $9\times10$ array of in-vessel optical fibres with toroidally viewing, horizontal lines of sight.\cite{cziegler-2010,cziegler-2012,zweben-rsi} The plasma emission collected in the views is filtered for He I ($587\,\text{nm}$) line emission that is locally enhanced in the object plane by an extended He gas puff from a nearby nozzle. Because the helium neutral density changes relatively slowly in space and time, rapid fluctuations in He I emission are caused by local plasma density and temperature fluctuations. The emission can be parameterized as proportional to $n_\text{e}^\alpha T_\text{e}^\beta$, with the exponents $\alpha$ and $\beta$ dependent upon the local electron density $n_\text{e}$ and temperature $T_\text{e}$.\cite{stotler,zweben-rsi} The optical fibres are coupled to high sensitivity avalanche photo diodes and the signals are digitized at a rate of $2\times10^6$ frames per second. The viewing area covers the major radius from $88.00$ to $91.08\,\text{cm}$ and vertical coordinate from $-4.51$ to $-1.08\,\text{cm}$ with an in-focus spot size of $3.8\,\text{mm}$ for each of the 90 individual channels. 
The limiter radius mapped to the GPI view position is at $R=91.0\,\text{cm}$.

Fluctuation statistics are here presented from the SOL of four Alcator C-Mod plasmas in different parameter regimes and confinement modes. Table~\ref{tab:plasma} gives the confinement mode, shot number and the duration of the time interval used for the following statistical analysis, during which all plasma parameters in the SOL are stationary. Also given in \Tabref{tab:plasma} are the axial magnetic field $B_0$, plasma current $\Ip$, the Greenwald fraction of the line-averaged density $\overline{n}_\text{e}$, and the ICRF heating power $P_\text{RF}$ during the time interval investigated. Here the Greenwald density is given by $n_\text{G}=(I_\text{p}/\pi a^2)10^{20}\,\text{m}^{-3}$, where $I_\text{p}$ is given in units of MA and $a$ is in units of meters.\cite{greenwaldlimit}

\begin{table}
\centering
\begin{tabular}{c|c|c|c|c|c|c}
Plasma state & Shot number & Duration$\,/\,\text{ms}$ & $B_0\,/\,\text{T}$ & $\Ip\,/\,\text{MA}$ & $\overline{n}_\text{e}/n_\text{G}$ & $P_\text{RF}\,/\,\text{MW}$ \\ 
\hline
Ohmic low density  & 1150618021 & 250 & 4.0 & 0.6 & 0.3 & 0   \\
Ohmic high density & 1150618036 & 460 & 4.0 & 0.6 & 0.6 & 0   \\
ELM-free H-mode   & 1110201011 & 100 & 5.4 & 1.2 & 0.5 & 3.0 \\
EDA H-mode         & 1110201016 & 225 & 5.4 & 0.9 & 0.6 & 3.0
\end{tabular}
\caption{List of plasma discharges giving the confinement state, shot number, duration of the time interval used for statistical analysis, axial magnetic field on axis, plasma current, Greenwald fraction of line-averaged core plasma density, and ICRF heating power.}
\label{tab:plasma}
\end{table}

The ohmically heated plasma states were part of a density scan study and the highest density case has a Greenwald density fraction of $\overline{n}_\text{e}/n_\text{G}=0.6$ and a fully detached divertor. These plasma states are included here also for the reason of comparison to results obtained from lower density ohmic plasmas previously published in Refs.~\onlinecite{garcia-acm-psi,garcia-acm-aps,theodorsen-nf}.
For the ELM-free H-mode case, there is only data time series of $100\,\text{ms}$ duration under stationary conditions in the SOL, implying that this case is not as well converged in the following statistical analysis as the other confinement states investigated here.

The measurement data for each diode view position is rescaled such as to have vanishing mean and unit standard deviation. Thus, a measured signal $\Phi(t)$ is normalized as $\wt{\Phi}=(\Phi-\overline{\Phi})/\overline{\Phi}_\text{rms}$, where $\overline{\Phi}$ and $\overline{\Phi}_\text{rms}$ are the moving average and standard deviation taken over a window of approximately $8\,\text{ms}$ duration in order to remove low-frequency trends in the signals.

\section{Fluctuation statistics}

\begin{figure}
\centering
\includegraphics[width=10cm]{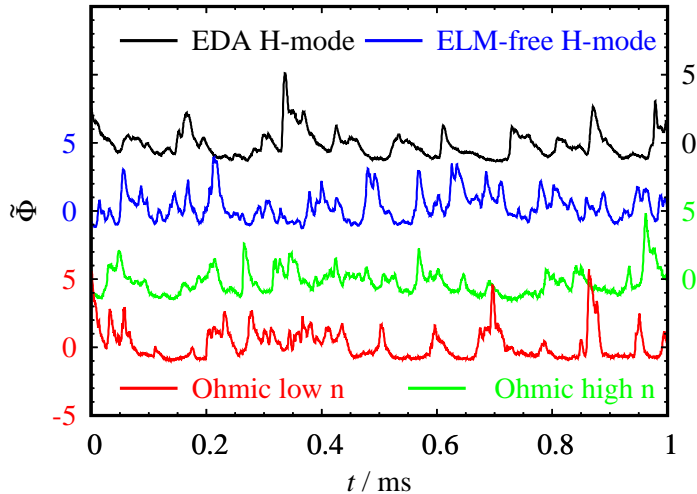}
\caption{Data time series recorded by the GPI diagnostic measured at $(R,Z)=(90.69,-2.99)\,\text{cm}$ for various plasma parameters and confinement modes. All time series are rescaled such as to have vanishing mean and unit standard deviation.}
\label{fig:traw}
\end{figure}

A short part of the detrended GPI data time series measured at $(R,Z)=(90.69,-2.99)\,\text{cm}$ for the four discharges listed in Tab.~\ref{tab:plasma} is presented in \Figref{fig:traw}, clearly showing the frequent occurrence of large-amplitude bursts in all confinement modes. The fluctuation data time series in the two H-mode cases appear qualitatively similar to those in the two ohmic plasma states. It should be noted that burst amplitudes several times the rms level occurs frequently in all cases.

\begin{figure}
\centering
\includegraphics[width=10cm]{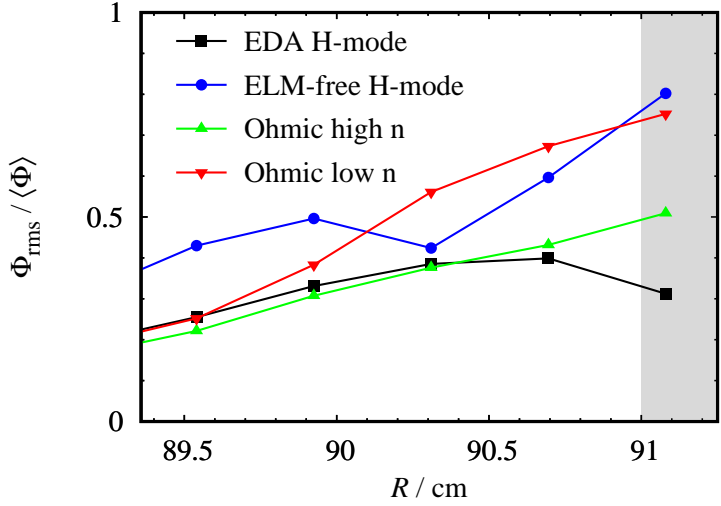}
\caption{Radial profile of the relative fluctuation level for the GPI intensity measured at $Z=-2.99\,\text{cm}$ for various plasma parameters and confinement modes. The shaded region indicates the limiter shadow region.}
\label{fig:rfl}
\end{figure}

The radial variation of the relative fluctuation level is presented in \Figref{fig:rfl}, showing an increase radially outwards in the \sol\ and order unity fluctuation levels in the far-\sol\ for all confinement modes. The high density ohmic case and the EDA H-mode case are indistinguishable except for a small difference in the limiter shadow region. Also the skewness and flatness moments increase with radial distance into the SOL. A varying degree of the injected neutral gas will be ionized in the SOL, depending on the electron density and temperature. The gas puff imaging diagnostic can therefore not be used to distinguish the absolute fluctuation amplitudes in any plasma parameter.\cite{stotler,zweben-rsi}

\begin{figure}
\centering
\includegraphics[width=10cm]{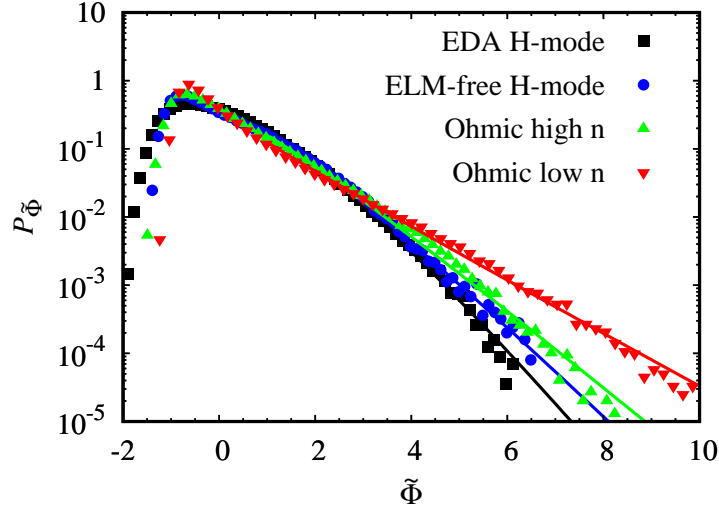}
\caption{Probability density function of the GPI intensity signals measured at $(R,Z)=(90.69,-2.99)\,\text{cm}$ for various plasma parameters and confinement modes. The full lines show the tails of Gamma distributions.}
\label{fig:pdf}
\end{figure}

The probability density functions (PDFs) for the detrended intensity fluctuations at $(R,Z)=(90.69,-2.99)\,\text{cm}$ are presented in \Figref{fig:pdf}. For all confinement modes there is a pronounced tail towards large fluctuation amplitudes, as expected from the frequent occurence of large-amplitude bursts in the underlying time series presented in \Figref{fig:traw}. These are attributed to the excess particles and heat in blob-like filaments propagating through the SOL. Also shown in \Figref{fig:pdf} are the predictions of a Gamma distribution for the large-amplitude tail of the PDFs. This is clearly a good description of both the ohmic and H-mode cases, with the strongest intermittency for the low density ohmic case. The shape parameter for the Gamma distributions, which is the ratio of the pulse duration and average waiting times, are $3/4$ for the low density ohmic case, $2$ for the high density ohmic case, $3$ for the ELM-free H-mode, and $5$ for the EDA H-mode.

\begin{figure}
\centering
\includegraphics[width=10cm]{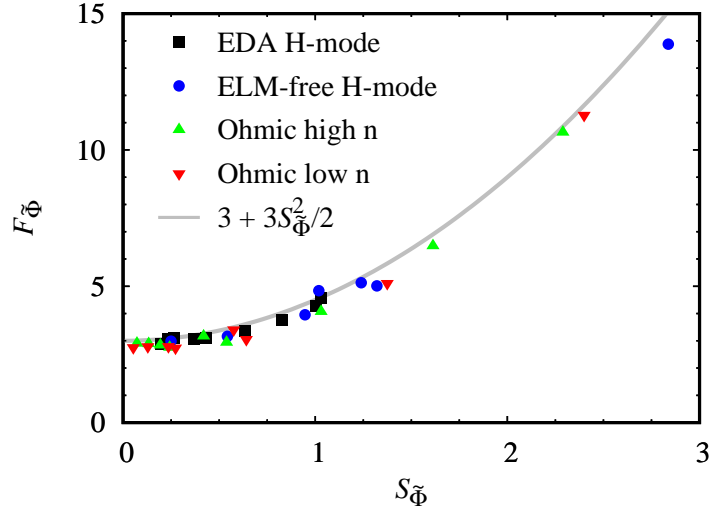}
\caption{Flatness versus skewness moments for GPI intensity signals measured in the SOL at $Z=-2.99\,\text{cm}$ for various plasma parameters and confinement modes.}
\label{fig:kvs}
\end{figure}

In order to further demonstrate the intermittency of the fluctuations, the sample flatness moment is plotted against the sample skewness moment in \Figref{fig:kvs} for all GPI diode view positions in the SOL region and all discharges listed in Tab.~\ref{tab:plasma}. The full line shows the parabolic relation between flatness and skewness predicted by a stochastic model describing the fluctuations as a super-position of uncorrelated exponential pulses with an exponential amplitude distribution.\cite{garcia-prl,kg,tgr,gktp} In agreement with the results in \Figsref{fig:rfl} and~\ref{fig:pdf}, the skewness and flatness moments both increase radially outwards in the scrape-off layer and their parabolic relation is in excellent agreement with predictions of the stochastic model.

\begin{figure}
\centering
\includegraphics[width=10cm]{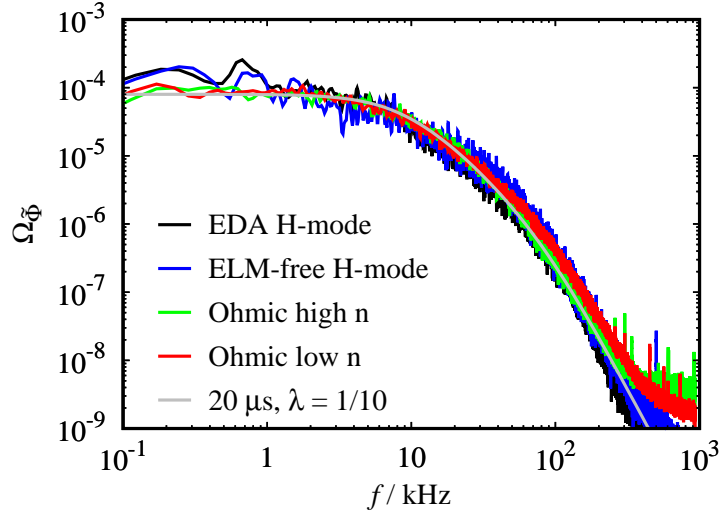}
\caption{Frequency power spectral density of GPI intensity signals measured at $(R,Z)=(90.69,-2.99)\,\text{cm}$ for various plasma parameters and confinement modes. Also shown is the spectrum predicted by the stochastic model.}
\label{fig:psd}
\end{figure}

\begin{figure}
\centering
\includegraphics[width=10cm]{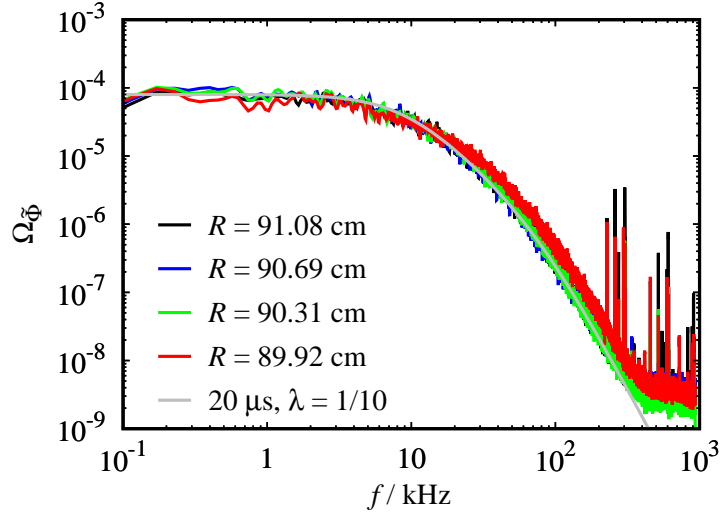}
\caption{Frequency power spectral density of GPI intensity signals measured at $Z=-2.99\,\text{cm}$ for various radial positions in the SOL for the high density ohmic plasma. Also shown is the spectrum predicted by the stochastic model.}
\label{fig:psdohmic}
\end{figure}

\begin{figure}
\centering
\includegraphics[width=10cm]{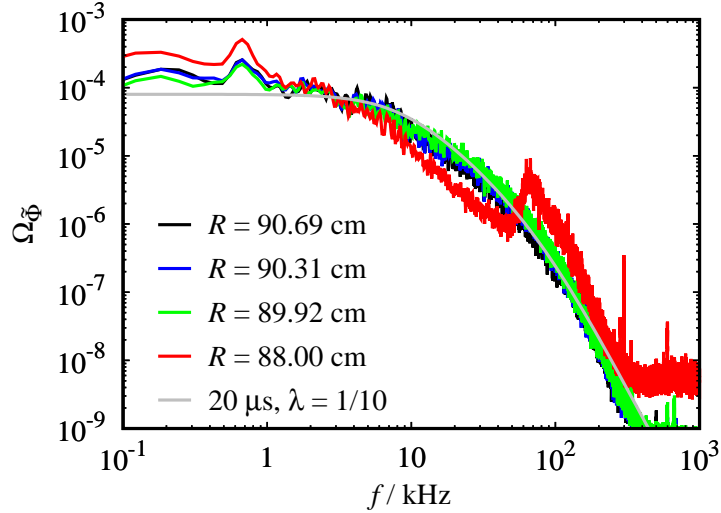}
\caption{Frequency power spectral density of GPI intensity signals measured at $Z=-2.99\,\text{cm}$ for various radial positions in the SOL for the EDA H-mode. Also shown is the spectrum predicted by the stochastic model.}
\label{fig:psdedah}
\end{figure}

The frequency power spectral densities $\Omega_{\wt{\Phi}}$ for the GPI fluctuation data time series measured at $(R,Z)=(90.69,-2.99)\,\text{cm}$ are presented in \Figref{fig:psd} as function of linear frequency $f$ for various plasma parameters and confinement modes. These are practically identical for the two ohmic and the two H-mode plasmas, and are well described the frequency spectrum predicted by a stochastic model describing the fluctuations as a super-position of uncorrelated exponential pulses with a duration time of $20\,\mu\text{s}$ and pulse asymmetry parameter (that is, pulse rise time over duration time) of $10^{-1}$.\cite{theodorsen-nf,gt} The shape of the frequency power spectrum is furthermore similar for all radial position in the SOL. One example of this is presented in \Figref{fig:psdohmic}, showing the spectra for various radial position in the SOL for the high density ohmic plasma case. The large spikes at high frequencies are due to measurement noise. Another example of this is presented in \Figref{fig:psdedah} for the EDA H-mode, which also shows a spectral peak at approximately $80\,\text{kHz}$ due to the QCM for the innermost GPI diode view position at $R=88.00\,\text{cm}$. Again, in the SOL the power spectral densities have the same shape for all radial positions and as for the ohmic plasmas.

In order to reveal the statistical properties of large-amplitude events in the time series, a standard conditional averaging technique is utilized. Events when the intensity signal is above a specified amplitude threshold value are recorded. The algorithm searches the signal for the largest amplitude events, and records conditional windows centred around the time of peak amplitude in the signal whenever the amplitude condition is satisfied. These sub-records are then averaged over all events to give the conditionally averaged wave-form associated with large-amplitude events in the signal. Overlap of conditional sub-records are avoided in order to ensure statistical independence of the events. Several hundred events are recorded for each of the time series investigated here.

\begin{figure}
\centering
\includegraphics[width=10cm]{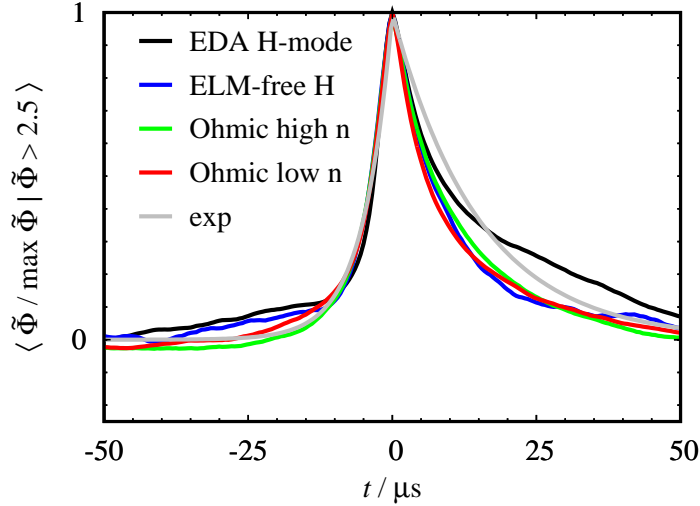}
\caption{Conditionally averaged wave form for large-amplitude events in the GPI intensity signals measured at $(R,Z)=(90.69,-2.99)\,\text{cm}$ for various plasma parameters and confinement modes. Also shown is a two-sided exponential pulse with a rise time of $5\,\mu\text{s}$ and fall time of $15\,\mu\text{s}$.}
\label{fig:cav}
\end{figure}

The conditionally averaged wave-forms calculated for the position $(R,Z)=(91.08,-2.99)\,\text{cm}$ are presented in \Figref{fig:cav} for a threshold value given by $2.5$ times the standard deviation. For all confinement modes there is on average a large peak nearly four times the standard deviation of the full time series. For the both ohmic and H-mode plasmas, the wave-form is well described by a two-sided exponential pulse shape with a rise time of $5\,\mu\text{s}$ and a fall time of approximately $15\,\mu\text{s}$. The somewhat longer pulse fall time for the EDA H-mode is likely due to the stronger degree of pulse overlap suggested by the relatively large shape parameter of the Gamma distribution presented in \Figref{fig:pdf}.

\begin{figure}
\centering
\includegraphics[width=10cm]{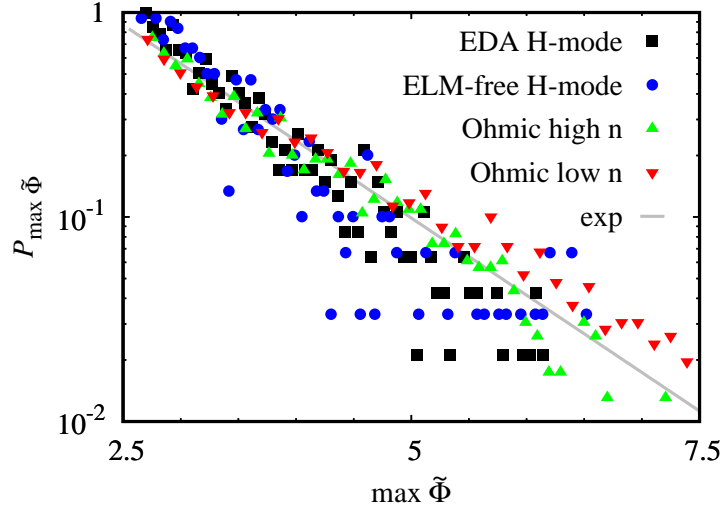}
\caption{PDF of peak amplitudes in the GPI intensity signals above $2.5$ standard deviations measured at $(R,Z)=(90.69,-2.99)\,\text{cm}$ for various plasma parameters and confinement modes. The full line shows a truncated exponential distribution.}
\label{fig:peak}
\end{figure}

For each crossing of the $2.5$ rms threshold level, the peak amplitudes are also recorded. Figure~\ref{fig:peak} shows the distribution of these peak amplitudes. The full line shows the prediction of a truncated exponential distribution with a mean amplitude of $3.65$, which is the same as the peak value for the conditionally averaged wave-form presented in \Figref{fig:cav}. Within the scatter due to finite duration of the time series, this is clearly an excellent description of the measured data. Similar results have previously been found for low density ohmic plasmas using both GPI and electric probe measurements.\cite{garcia-acm-psi,garcia-acm-aps,theodorsen-nf,ghp,kube-ppcf,theodorsen-ppcf,garcia-nme}

\begin{figure}
\centering
\includegraphics[width=10cm]{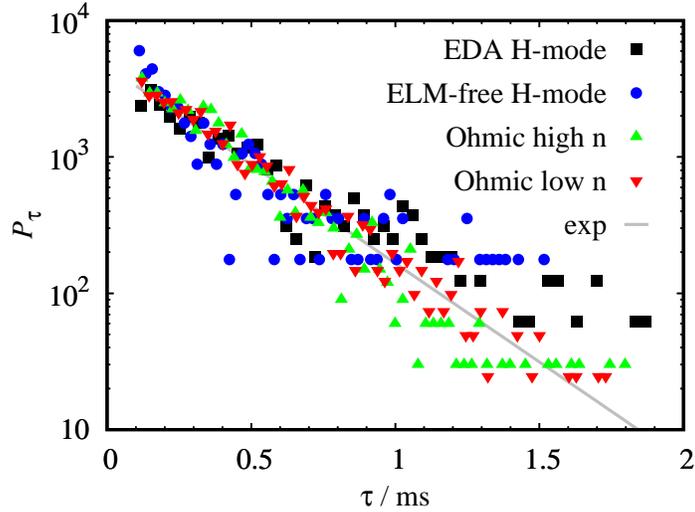}
\caption{PDF of waiting times between large-amplitude events with peak amplitudes above $2.5$ standard deviations measured at $(R,Z)=(90.69,-2.99)\,\text{cm}$ for various plasma parameters and confinement modes. The full line shows a truncated exponential distribution.}
\label{fig:wait}
\end{figure}

The PDF for the waiting times between large-amplitude bursts in the GPI intensity signals measured at $(R,Z)=(91.08,-2.99)\,\text{cm}$ are presented in \Figref{fig:wait}, again for a threshold value given by $2.5$ times the standard deviation. For the ohmic and H-mode plasmas, this is well described by an exponential distribution with an average waiting time of roughly $0.3\,\text{ms}$. Such an exponential distribution of waiting times is consistent with a process with uncorrelated events.

\section{Discussion and conclusions}

The radial motion of plasma filaments containing excess particles and heat leads to strongly intermittent fluctuations in the SOL of magnetically confined plasmas, and may lead to enhanced levels of plasmas--wall interactions that is a serious issue for the next generation, high duty cycle confinement experiments and future fusion reactors. In this work, it is for the first time demonstrated that the statistical properties of these fluctuations are the same in L- and H-mode plasmas. The fluctuations in the SOL are not influenced by the presence of a transport barrier in the edge region nor by the presence of mode structure such as the QCM in Alcator C-Mod.

This suggests the presence of universal statistical properties of the SOL fluctuations. In particular, the average large-amplitude fluctuation wave form is well described by an exponential function, and the peak amplitudes of the fluctuations as well as the waiting times between them are exponentially distributed in both ohmic and H-mode plasmas. This is evidence that supports a stochastic model describing the fluctuations as a super-position of uncorrelated pulses. This model predicts a Gamma distribution of the fluctuations where the shape parameter is given by the ratio of the pulse duration and waiting times. This is in excellent agreement with GPI measurement data from Alcator C-Mod, comprising a range of line-averaged densities and various confinement modes. This complements previous investigations of a set of low density ohmic plasmas.\cite{garcia-acm-psi,garcia-acm-aps,theodorsen-nf}

The model further predicts a frequency power spectral density that is independent of the degree of pulse overlap and the amplitude distribution of the pulses. Hence, the power spectrum is expected to be self-similar for all radial positions in the SOL. This is indeed shown to be the case for both ohmic and H-mode plasmas. This suggests that both the near- and the far-SOL fluctuations are due to uncorrelated exponential pulses but with much more pulse overlap close to the separatrix. These observations run contrary to the ideas that the shape of the power spectrum arises from the interaction of turbulent eddies or self-similar processes. The fluctuation statistics are shown to be the same in both Ohmic plasmas and high confinement modes. The pulse overlap is observed to decrease with radius into the SOL. This is likely due to significant poloidal and toroidal flows in the edge region as well as dispersion of the blob-like filaments as they propagate. It is to be noted that in the framework of the stochastic model, all the plasma in the SOL is due to radial motion of filament structures. This results in broad plasma profiles and enhanced levels of plasma--wall interactions.\cite{militello-nf-2016,militello-ppcf-2016,walkden-ppcf-2017,garcia-prl,kg,tg,gktp,gt,tgr,tg-pdf,tg-x,marandet-2016,marandet-2017}

\section{Acknowledgements}

This work was supported with financial subvention from the Research Council of Norway under grant 240510/F20 and the U.S. Department of Energy, Office of Science, Office of Fusion Energy Sciences, using User Facility Alcator C-Mod, under Award Number DE-FC02-99ER54512-CMOD. O.~E.~G., R.~K.\ and A.~T.\ acknowledge the generous hospitality of the MIT Plasma Science and Fusion Center where this work was conducted.


\begin{thebibliography}{99}
%
%
\bibitem{labombard-2001}
B.~La{B}ombard, R.~L.~Boivin, M.~Greenwald, J.~Hughes, B.~Lipschultz, D.~Mossessian, C.~S.~Pitcher, J.~L.~Terry, S.~J.~Zweben, and Alcator Group, \PP\ {\bf 8}, 2107 (2001).
%
\bibitem{pigarov}
A.~Yu.~Pigarov, S.~I.~Krasheninnikov, T.~D.~Rognlien, M.~J.~Schaffer, and W.~P.~West, \PP\ {\bf 9}, 1287 (2002).
%
\bibitem{pitts}
R.~A.~Pitts, J.~P.~Coad, D.~P.~Coster, G.~Federici, W.~Fundamenski, J.~Horacek, K.~Krieger, A.~Kukushkin, J.~Likonen, G.~F.~Matthews, M.~Rubel, J.~D.~Strachan, and JET-EFDA contributors, \PPCF\d {\bf 47}, B303 (2005).
%
\bibitem{whyte}
D.~G.~Whyte, B.~L.~Lipschultz, P.~C.~Stangeby, J.~Boedo, D.~L.~Rudakov, J.~G.~Watkins, and W.~P.~West, \PPCF\ {\bf 47}, 1579 (2005).
%
\bibitem{lipschultz}
B.~Lipschultz, X.~Bonnin, G.~Counsell, A.~Kallenbach, A.~Kukushkin, K.~Krieger, A.~Leonard, A.~Loarte, R.~Neu, R.~A.~Pitts, T.~Rognlien, J.~Roth, C.~Skinner, J.~L.~Terry, E.~Tsitrone, D.~Whyte, S.~Zweben, N.~Asakura, D.~Coster, R.~Doerner, R.~Dux, G.~Federici, M.~Fenstermacher, W.~Fundamenski, P.~Ghendrih, A.~Herrmann, J.~Hu, S.~Krasheninnikov, G.~Kirnev, A.~Kreter, V.~Kurnaev, B.~LaBombard, S.~Lisgo, T.~Nakano, N.~Ohno, H.~D.~Pacher, J.~Paley, Y.~Pan, G.~Pautasso, V.~Philipps, V.~Rohde, D.~Rudakov, P.~Stangeby, S.~Takamura, T.~Tanabe, Y.~Yang, and S.~Zhu, \NF\ {\bf 47}, 1189 (2007).
%
\bibitem{dm-pwi}
D.~A.~D'{I}ppolito and J.~R.~Myra, \PP\ {\bf 15}, 082316 (2008).
%
\bibitem{brooks}
J.~N.~Brooks, J.~P.~Allain, R.~P.~Doerner, A.~Hassanein, R.~Nygren, T.~D.~Rognlien, and D.~G.~Whyte, \NF\ {\bf 49}, 035007 (2009).
%
\bibitem{marandet}
Y.~Marandet, A.~Mekkaoui, D.~Reiter, P.~B{\"o}rner, P.~Genesio, F.~Catoire, J.~Rosato, H.~Capes, L.~Godbert-Mouret, M.~Koubiti, and R.~Stamm, \NF\ {\bf 51}, 083035 (2011).
%
\bibitem{dmz}
D.~A.~D'{I}ppolito, J.~R.~Myra, and S.~J.~Zweben, \PP\ {\bf 18}, 060501 (2011).
%
\bibitem{birkenmeier}
G.~Birkenmeier, P.~Manz, D.~Carralero, F.~M.~Laggner, G.~Fuchert, K.~Krieger, H.~Maier, F.~Reimold, K.~Schmid, R.~Dux, T.~P{\"u}tterich, M.~Willensdorfer, E.~Wolfrum, and The ASDEX Upgrade Team, \NF\ {\bf 55}, 033018 (2015).
%
%
\bibitem{boedo1}
J.~A.~Boedo, D.~Rudakov, R.~Moyer, S.~Krasheninnikov, D.~Whyte, G.~McKee, G.~Tynan, M.~Schaffer, P.~Stangeby, P.~West, S.~Allen, T.~Evans, R.~Fonck, E.~Hollmann, A.~Leonard, A.~Mahdavi, G.~Porter, M.~Tillack and G.~Antar, \PP\ {\bf 8}, 4826 (2001).
%
\bibitem{rudakov}
D.~L.~Rudakov, J.~A.~Boedo, R.~A.~Moyer, S.~Krasheninnikov, A.~W.~Leonard, M.~A.~Mahdavi, G.~R.~Mc{K}ee, G.~D.~Porter, P.~C.~Stangeby, J.~G.~Watkins, W.~P.~West, D.~G.~Whyte, and G.~Antar, \PPCF\ {\bf 44}, 717 (2002).
%
\bibitem{antar}
G.~Y.~Antar, G.~Counsell, Y.~Yu, B.~La{B}ombard, and P.~Devynck, \PP\ {\bf 10}, 419 (2003).
%
\bibitem{boedo2}
J.~A.~Boedo, D.~L.~Rudakov, R.~A.~Moyer, G.~R.~Mc{K}ee, R.~J.~Colchin,
M.~J.~Schaffer, P.~G.~Stangeby, W.~P.~West, S.~L.~Allen, T.~E.~Evans, R.~J.~Fonck,
E.~M.~Hollmann, S.~Krasheninnikov, A.~W.~Leonard, W.~Nevins, M.~A.~Mahdavi,
G.~D.~Porter, G.~R.~Tynan, D.~G.~Whyte and X.~Xu, \PP\ {\bf 10}, 1670 (2003).
%
\bibitem{carreras}
B.~A.~Carreras, \JNM\ {\bf 337-339}, 315 (2005).
%
\bibitem{graves}
J.~P.~Graves, J.~Horacek, R.~A.~Pitts and K.~I.~Hopcraft, \PPCF\ {\bf 47},  L1 (2005).
%
\bibitem{horacek}
J.~Horacek, R.~A.~Pitts and J.~P.~Graves, \CJP\ {\bf 55}, 271 (2005).
%
\bibitem{rudakov-nf}
D.~L.~Rudakov, J.~A.~Boedo, R.~A.~Moyer, P.~C.~Stangeby, J.~G.~Watkins, D.~G.~Whyte, L.~Zeng, N.~H.~Brooks, R.~P.~Doerner, T.~E.~Evans, M.~E.~Fenstermacher, M.~Groth, E.~M.~Hollmann, S.~I.~Krasheninnikov, C.~J.~Lasnier, A.~W.~Leonard, M.~A.~Mahdavi, G.~R.~McKee, A.~G.~Mc{L}ean, A.~Yu.~Pigarov, W.~R.~Wampler, G.~Wang, W.~P.~West and C.~P.~C.~Wong, \NF\ {\bf 45}, 1589 (2005).
%
\bibitem{garcia-tcv-fec}
O.~E.~Garcia, J.~Horacek, R.~A.~Pitts, A.~H.~Nielsen, W.~Fundamenski, V.~Naulin, and J.~Juul Rasmussen, \NF\ {\bf 47}, 667 (2007). 
%
\bibitem{garcia-tcv-eps}
O.~E.~Garcia, R.~A.~Pitts, J.~Horacek, J.~Madsen, V.~Naulin, A.~H.~Nielsen, and J.~Juul Rasmussen, \PPCF\ {\bf 49}, B47 (2007).
%
\bibitem{xu}
G.~S.~Xu, V.~Naulin, W.~Fundamenski, C.~Hidalgo, J.~A.~Alonso, C.~Silva, B.~Goncalves, A.~H.~Nielsen, J.~Juul Rasmussen, S.~I.~Krasheninnikov, B.~N.~Wan, M.~Stamp and JET EFDA Contributors, \NF\ {\bf 49}, 092002 (2009).
%
\bibitem{garcia-pfr}
O.~E.~Garcia, \PFR\ {\bf 4}, 019 (2009).
%
\bibitem{horacek-asdex}
J.~Horacek, J.~Adamek, H.~W.~Mu{\"u}ller, J.~Seidl, A.~H.~Nielsen, V.~Rohde, F.~Mehlmann, C.~Ionita, E.~Havl{\'i}{\v{c}}kov{\'a} and the ASDEX Upgrade Team, \NF\ {\bf 50}, 105001 (2010).
%
\bibitem{militello}
F.~Militello, P.~Tamain, W.~Fundamenski, A.~Kirk, V.~Naulin, A.~H.~Nielsen and the MAST team, \PPCF\ {\bf 55}, 025005 (2013).
%
\bibitem{carralero-nf}
D.~Carralero, G.~Birkenmeier, H.~W.~M{\"u}ller, P.~Manz, P.~de{M}arne, S.~H.~M{\"u}ller, F.~Reimold, U.~Stroth, M.~Wischmeier, E.~Wolfrum and {T}he ASDEX {U}pgrade {T}eam, \NF\ {\bf 54}, 123005 (2014).
%
\bibitem{zweben-2015}
S.~J.~Zweben, W.~M.~Davis, S.~M.~Kaye, J.~R.~Myra, R.~E.~Bell, B.~P.~Le{B}lanc, R.~J.~Maqueda, T.~Munsat, S.~A.~Sabbagh, Y.~Sechrest, D.~P.~Stotler and the NSTX Team, \NF\ {\bf 55}, 093035 (2015).
%
\bibitem{carralero-prl}
D.~Carralero, P.~Manz, L.~Aho-Mantila, G.~Birkenmeier, M.~Brix, M.~Groth, H.~W.~M{\"u}ller, U.~Stroth, N.~Vianello, E.~Wolfrum, ASDEX Upgrade team, JET Contributors, and EUROfusion MST1 Team, \PRL\ {\bf 115}, 215002 (2015).
%
\bibitem{zweben-2016}
S.~J.~Zweben, J.~R.~Myra, W.~M.~Davis, D.~A.~D’{I}ppolito, T.~K.~Gray, S.~M.~Kaye, B.~P.~Le{B}lanc, R.~J.~Maqueda, D.~A.~Russell, D.~P.~Stotler and the NSTX-U Team, \PPCF\ {\bf 58}, 044007 (2016).
%
\bibitem{boedo3}
J.~A.~Boedo, J.~R.~Myra, S.~Zweben, R.~Maingi, R.~J.~Maqueda, V.~A.~Soukhanovskii, J.~W.~Ahn, J.~Canik, N.~Crocker, D.~A.~D'{I}ppolito, R.~Bell, H.~Kugel, B.~Leblanc, L.~A.~Roquemore, D.~L.~Rudakov and NSTX Team, \PP\ {\bf 21}, 042309 (2014).
%
\bibitem{walkden}
N.~R.~Walkden, A.~Wynn, F.~Militello, B.~Lipschultz, G.~Matthews, C.~Guillemaut, J.~Harrison, D.~Moulton and JET Contributors, \NF\ {\bf 57}, 036016 (2017).
%
%
\bibitem{garcia-acm-psi}
O.~E.~Garcia, I.~Cziegler, R.~Kube, B.~La{B}ombard, and J.~L.~Terry, \JNM\ {\bf 438}, S180 (2013).
%
\bibitem{garcia-acm-aps}
O.~E.~Garcia, S.~M.~Fritzner, R.~Kube, I.~Cziegler, B.~La{B}ombard, and J.~L.~Terry, \PP\ {\bf 20}, 055901 (2013).
%
\bibitem{theodorsen-nf}
A.~Theodorsen, O.~E.~Garcia, R.~Kube, B.~La{B}ombard, and J.~L.~Terry, \NF\ {\bf 57}, 114004 (2017).
%
\bibitem{ghp}
O.~E.~Garcia, J.~Horacek, and R.~A.~Pitts, \NF\ {\bf 55}, 062002 (2015).
%
\bibitem{theodorsen-ppcf}
A.~Theodorsen, O.~E.~Garcia, J.~Horacek, R.~Kube, and R.~A.~Pitts, \PPCF\ {\bf 58}, 044006 (2016).
%
\bibitem{kube-ppcf}
R.~Kube, A.~Theodorsen, O.~E.~Garcia, B.~La{B}ombard, and J.~L.~Terry, \PPCF\ {\bf 58}, 054001 (2016).
%
\bibitem{garcia-nme}
O.~E.~Garcia, R.~Kube, A.~Theodorsen, J.-G.~Bak, S.-H.~Hong, H.-S.~Kim, the KSTAR Project Team, and R.~A.~Pitts, \NME\ {\bf 12}, 36 (2017).
%
\bibitem{militello-nf-2016}
F.~Militello and J.~T.~Omotani, \NF\ {\bf 56}, 104004 (2016).
%
\bibitem{militello-ppcf-2016}
F.~Militello and J.~T.~Omotani, \PPCF\ {\bf 58}, 125004 (2016).
%
\bibitem{walkden-ppcf-2017}
N.~R.~Walkden, A.~Wynn, F.~Militello, B.~Lipschultz, G.~Matthews, C.~Guillemaut, J.~Harrison, D.~Moulton, and JET Contributors, \PPCF\ {\bf 59}, 085009 (2017).
%
%
\bibitem{gktp}
O.~E.~Garcia, R.~Kube, A.~Theodorsen, and H.~L.~P{\'e}cseli, \PP\ {\bf 23}, 052308 (2016).
%
\bibitem{garcia-prl}
O.~E.~Garcia, \PRL\ {\bf 108}, 265001 (2012).
%
\bibitem{kg}
R.~Kube and O.~E.~Garcia, \PP\ {\bf 22}, 012502 (2015).
%
\bibitem{tgr}
A.~Theodorsen, O.~E.~Garcia, and M.~W.~Rypdal, \PS\ {\bf 92}, 054002 (2017).
%
\bibitem{tg}
A.~Theodorsen and O.~E.~Garcia, \PP\ {\bf 23}, 040702 (2016).
%
\bibitem{gt}
O.~E.~Garcia and A.~Theodorsen, \PP\ {\bf 24}, 032309 (2017).
%
\bibitem{tg-pdf}
A.~Theodorsen and O.~E.~Garcia, ``Probability distribution functions for intermittent scrape-off layer plasma fluctuations'' accepted for publication in \PPCF.
%
\bibitem{tg-x}
A.~Theodorsen and O.~E.~Garcia, ``Level crossings and excess times due to a super-position of uncorrelated exponential pulses'' submitted to \PRE.
%
%
\bibitem{marandet-2016}
Y.~Marandet, N.~Nace, M.~Valentinuzzi, P.~Tamain, H.~Bufferand, G.~Ciraolo, P.~Genesio, and N Mellet, \PPCF\ {\bf 58}, 114001 (2016).
%
\bibitem{marandet-2017}
Y.~Marandet, H.~Bufferand, N.~Nace, M.~Valentinuzzi, G.~Ciraolo, P.~Tamain J.~Bucalossi, D.~Galassi, Ph.~Ghendrih, N.~Mellet, and E.~Serre, \NME\ {\bf 12}, 931 (2017).
%
%
\bibitem{hutchinson}
I.~H.~Hutchinson, R.~Boivin, F.~Bombarda, P.~Bonoli, S.~Fairfax, C.~Fiore, J.~Goetz, S.~Golovato, R.~Granetz, M.~Greenwald, S.~Horne, A.~Hubbard, J.~Irby, B.~La{B}ombard, B.~Lipschultz, E.~Marmar, G.~Mc{C}racken, M.~Porkolab, J.~Rice, J.~Snipes, Y.~Takase, J.~Terry, S.~Wolfe, C.~Christensen, D.~Garnier, M.~Graf, T.~Hsu, T.~Luke, M.~May, A.~Niemczewski, G.~Tinios, J.~Schachter, and J.~Urbahn, \PP\ {\bf 1}, 1511 (1994).
%
\bibitem{greenwald-2013}
M.~Greenwald, A.~Bader, S.~Baek, H.~Barnard, W.~Beck, W.~Bergerson, I.~Bespamyatnov, M.~Bitter, P.~Bonoli, M.~Brookman, D.~Brower, D.~Brunner, W.~Burke, J.~Candy, M.~Chilenski, M.~Chung, M.~Churchill, I.~Cziegler, E.~Davis, G.~Dekow, L.~Delgado-Aparicio, A.~Diallo, W.~Ding, A.~Dominguez, R.~Ellis, P.~Ennever, D.~Ernst, I.~Faust, C.~Fiore, E.~Fitzgerald, T.~Fredian, O.~E.~Garcia, C.~Gao, M.~Garrett, T.~Golfinopoulos, R.~Granetz, R.~Groebner, S.~Harrison, R.~Harvey, Z.~Hartwig, K.~Hill, J.~Hillairet, N.~Howard, A.~E.~Hubbard, J.~W.~Hughes, I.~Hutchinson, J.~Irby, A.~N.~James, A.~Kanojia, C.~Kasten, J.~Kesner, C.~Kessel, R.~Kube, B.~La{B}ombard, C.~ Lau, J.~Lee, K.~Liao, Y.~Lin, B.~Lipschultz, Y.~Ma, E.~Marmar, P.~Mc{G}ibbon, O.~Meneghini, D.~Mikkelsen, D.~Miller, R.~Mumgaard, R.~Murray, R.~Ochoukov, G.~Olynyk, D.~Pace, S.~Park, R.~Parker, Y.~Podpaly, M.~Porkolab, M.~Preynas, I.~Pusztai, M.~Reinke, J.~Rice, W.~Rowan, S.~Scott, S.~Shiraiwa, J.~Sierchio, P.~Snyder, B.~Sorbom, V.~Soukhanovskii, J.~Stillerman, L.~Sugiyama, C.~Sung, D.~Terry, J.~Terry, C.~Theiler, N.~Tsujii, R.~Vieira, J.~Walk, G.~Wallace, A.~White, D.~Whyte, J.~Wilson, S.~Wolfe, K.~Woller, G.~Wright, J.~Wright, S.~Wukitch, G.~Wurden, P.~Xu, C.~Yang, and S.~Zweben, \NF\ {\bf 53}, 104004 (2013).
%
\bibitem{greenwald-2014}
M.~Greenwald, A.~Bader, S.~Baek, M.~Bakhtiari, H.~Barnard, W.~Beck, W.~Bergerson, I.~Bespamyatnov, P.~Bonoli, D.~Brower, D.~Brunner, W.~Burke, J.~Candy, M.~Churchill, I.~Cziegler, A.~Diallo, A.~Dominguez, B.~Duval, E.~Edlund, P.~Ennever, D.~Ernst, I.~Faust, C.~Fiore, T.~Fredian, O.~Garcia, C.~Gao, J.~Goetz, T.~Golfinopoulos, R.~Granetz, O.~Grulke, Z.~Hartwig, S.~Horne, N.~Howard, A.~Hubbard, J.~Hughes, I.~Hutchinson, J.~Irby, V.~Izzo, C.~Kessel, B.~La{B}ombard, C.~Lau, C.~Li, Y.~Lin, B.~Lipschultz, A.~Loarte, E.~Marmar, A.~Mazurenko, G.~Mc{C}racken, R.~McDermott, O.~Meneghini, D.~Mikkelsen, D.~Mossessian, R.~Mumgaard, J.~Myra, E.~Nelson-Melby, R.~Ochoukov, G.~Olynyk, R.~Parker, S.~Pitcher, Y.~Podpaly, M.~Porkolab, M.~Reinke, J.~Rice, W.~Rowan, A.~Schmidt, S.~Scott, S.~Shiraiwa, J.~Sierchio, N.~Smick, J.~A.~Snipes, P.~Snyder, B.~Sorbom, J.~Stillerman, C.~Sung, Y.~Takase, V.~Tang, J.~Terry, D.~Terry, C.~Theiler, A.~Tronchin-{J}ames, N.~Tsujii, R.~Vieira, J.~Walk, G.~Wallace, A.~White, D.~Whyte, J.~Wilson, S.~Wolfe, G.~Wright, J.~Wright, S.~Wukitch, and S.~Zweben, \PP\ {\bf 21}, 110501 (2014).
%
%
\bibitem{labombard-qcm}
B.~La{B}ombard, T.~Golfinopoulos, J.~L.~Terry, D.~Brunner, E.~Davis, M.~Greenwald, J.~W.~Hughes, and Alcator C-Mod Team, \PP\ {\bf 21}, 056108 (2014).
%
\bibitem{terry-2005}
J.~L.~Terry, N.~P.~Basse, I.~Cziegler, M.~Greenwald, O.~Grulke, B.~La{B}ombard, S.~J.~Zweben, E.~M.~Edlund, J.~W.~Hughes, L.~Lin, Y.~Lin, M.~Porkolab, M.~Sampsell, B.~Veto, and S.~J.~Wukitch, \NF\ {\bf 45}, 1321 (2005).
%
\bibitem{cziegler-2010}
I.~Cziegler, J.~L.~Terry, J.~W.~Hughes, and B.~La{B}ombard, \PP\ {\bf 17}, 056120 (2010).
%
%
\bibitem{cziegler-2012}
I.~Cziegler, J.~L.~Terry, S.~J.~Wukitch, M.~L.~Garrett, C.~Lau, and Y.~Lin, \PPCF\ {\bf 54}, 105019 (2012).
%
\bibitem{zweben-rsi}
S.~J.~Zweben, J.~L.~Terry, D.~P.~Stotler, and R.~J.~Maqueda, \RSI\ {\bf 88}, 041101 (2017).
%
\bibitem{stotler}
D.~P.~Stotler, B.~La{B}ombard, J.~L.~Terry, and S.~J.~Zweben, \JNM\ {\bf 313-316}, 1066 (2003).
%
%
\bibitem{greenwaldlimit}
M.~Greenwald, \PPCF\ {\bf 44}, R27 (2002).
%
%
%
\end{thebibliography}
\end{document}